# Signatures of arithmetic simplicity in metabolic network architecture

William J Riehl[1], Paul L. Krapivsky[2], Sidney Redner[2] and Daniel Segrè[1,3]

[1]Program in Bioinformatics and Systems Biology, [2]Department of Physics, [3]Department of Biology, Department of Biomedical Engineering, Boston University, Boston, Massachusetts 02215.

**Abstract**

Metabolic networks perform some of the most fundamental functions in living cells, including energy transduction and building block biosynthesis. While these are the best characterized networks in living systems, understanding their evolutionary history and complex wiring constitutes one of the most fascinating open questions in biology, intimately related to the enigma of life's origin itself. Is the evolution of metabolism subject to general principles, beyond the unpredictable accumulation of multiple historical accidents? Here we search for such principles by applying to an artificial chemical universe some of the methodologies developed for the study of genome scale models of cellular metabolism. In particular, we use metabolic flux constraint-based models to exhaustively search for artificial chemistry pathways that can optimally perform an array of elementary metabolic functions. Despite the simplicity of the model employed, we find that the ensuing pathways display a surprisingly rich set of properties, including the existence of autocatalytic cycles and hierarchical modules, the appearance of universally preferable metabolites and reactions, and a logarithmic trend of pathway length as a function of input/output molecule size. Some of these properties can be derived analytically, borrowing methods previously used in cryptography. In addition, by mapping biochemical networks onto a simplified carbon atom reaction backbone, we find that several of the properties predicted by the artificial chemistry model hold for real metabolic networks. These findings suggest that optimality principles and arithmetic simplicity might lie beneath some aspects of biochemical complexity.

**Author Summary**

An open question in biology is whether the evolution of metabolic networks has been guided by general optimality principles beyond the unpredictable accumulation of historical accidents. Here we search for signatures of such optimality in an idealized artificial chemistry model, where it is feasible to systematically explore a complete set of efficient metabolic pathways of minimal length between any two compounds. These pathways display a modular organization of recurring topologies, including autocatalytic cycles, and a logarithmic dependence of pathway length on input/output molecule size. Across all pathways, we predict the emergence of ubiquitous metabolites, and a broad spectrum of reaction utilization, with certain reactions serving as universal steps. Similar properties hold for real metabolic networks, suggesting that optimality principles and arithmetic simplicity may underlie some aspects of biochemical complexity.

**Introduction**

The prominent role of metabolism in any biological process and the fact that a large portion of the environmental factors shaping living systems are ultimately metabolic in nature, suggest that strong selective forces have been acting on metabolic networks throughout the history of life. In laboratory evolution experiments [1-3] one can witness mostly short term metabolic adaptations, affecting metabolic enzyme regulation and fine tuning of kinetic parameters. However, especially during major transitions, such as the early stages of life's appearance or the rise of oxygen in the Earth's atmosphere, selective forces must have shaped the metabolic wiring itself [4]. Comparative genomics can provide top-down insight into some long-term evolution of metabolic pathways [5, 6]. In addition, studies of prebiotic chemistry scenarios have suggested possible seeds of biochemical organization from a bottom-up perspective [7, 8]. Yet, whether the long term evolution of metabolism was dominated by unpredictable frozen accidents, or by inevitable network optimization processes, remains a fundamental open question.

In a 1961 review, Baldwin and Krebs suggested that biochemical network topologies may reflect the adaptation toward optimally efficient metabolic strategies, and that manifold use of certain molecules may be a crucial element of this adaptation, as "it is indeed a general principle of evolution that multiple use is made of given resources."[9]. Some computational studies have proposed that the topology of specific metabolic pathways may have evolved towards maximal efficiency [10], minimal number of steps [11], or that network properties may reflect optimal organization [12]. Here we seek to address this problem by exploring a system that can reach a level of complexity comparable to the one observed in the union of all known metabolic pathways, yet is simple enough to allow efficient computation and analytical calculations. In addition, we wish to explore at an ecosystem-level the potential role of "metabolic multi-tasking", as suggested by Baldwin and Krebs. The increasing evidence of abundant horizontal gene transfer in the history of life suggests that this question may be indeed especially relevant at the ecosystem level, where the interchange of genetic information might have created a free economy of enzymes among simple organisms, allowing for the emergence of species that share common molecular tools [13]. Recent metagenomic studies of microbial consortia [14] also suggest the question of whether metabolic functions, more than individual species distributions, might be directly dependent on environmental conditions. Hence it is possible that hallmarks of metabolic optimality in metabolic network wiring may be observable at the level of global (multi-species) metabolic network [15, 16], more than at the individual species level. Do specific molecules, reactions or pathway topologies appear to be universally useful in a biochemical network, i.e., relevant for maximally efficient completion of several possible metabolic tasks, possibly across multiple organisms?

In the present work, we combine the study of an extremely simple artificial chemistry [17-20] with recent systems biology approaches [21, 22] to systematically compute pathways that are optimal for an array of elementary metabolic tasks, converting an input molecule into an output one. Behind the apparent complexity of the ensuing pathways, we identify recurring, modularly organized categories of network topologies, and analytically predictable trends in pathway length. In addition, we observe the emergence of "universal metabolic tools" across all optimal pathways. Finally, despite the huge gap in the underlying chemical rules, we find that some properties of real metabolic pathways are consistent with the patterns detected in the model, suggesting that fundamental optimality principles may have played a role in shaping biochemical networks.

**Results**

Our artificial chemistry consists of a set of $N$ possible molecules $\{a_1, a_2, a_3, ..., a_N\}$, that can participate in reversible ligation/cleavage reactions of the form $a_i + a_j \leftrightarrow a_k$, with $i + j = k$. This model could be viewed as the simplest possible string-based artificial chemistry [17]. The reaction network $R_N$ that includes all metabolites up to length $N$ and all possible reactions (of the order of $0.25N^2$) between them (Fig. 1) can be thought of as the underlying chemistry based on which specialized metabolic tasks could emerge. Here we were concerned with pathways, within the $R_N$ network, that can optimally perform a given metabolic task. In particular, we searched for optimal solutions to the problem of producing a specific end-product (e.g., $a_j$, with output flux $v_{out}$) from a single available nutrient (e.g., $a_i$, with input flux $v_{in}$). We define an optimal pathway as one that satisfies the following conditions: (i) it allows a steady state solution, i.e., a mass-conserving flow from input to output; (ii) it has maximal yield, and no waste [23], such that $v_{out}=v_{in}\cdot j/i$; and (iii) it has the fewest reaction steps possible. A pathway satisfying these conditions is termed a *minimal balanced pathway* (MBP) between $a_i$ and $a_j$, and will be denoted $a_i \Rightarrow a_j$. MBPs (also referred to below as optimal pathways) can be thought of as the pathways that are most efficient for a specific metabolic task, in the sense that they require the smallest possible number of different enzymes for producing the maximal possible yield [10, 24, 25].

Despite the simplicity of our artificial chemistry, identifying the MBPs between all possible input-output pairs in a given artificial chemistry $R_N$ is a challenge for large $N$. We implement three algorithms to approach this problem: a mixed integer linear programming (MILP) akin to flux balance analysis (FBA) [26]; an algorithm that uses enumeration of elementary flux modes [21]; and finally an iterative algorithm that gradually assembles new MBPs from already identified simple ones (see Methods). The three algorithms differ mainly in their scalability, and in their capacity to predict multiple degenerate solutions (see Table S1). A partial overview of the results of our calculations is shown in Fig. 2A and Fig. S1 (see Table S2 for a comprehensive list of MBPs).

Behind the apparent complexity of the topologies encountered in each of the different pathways, it is possible to observe the recurrence of three fundamental categories: each MBP functions either as a pure "addition chain" [27], where smaller metabolites are progressively added together to build the target molecule, or as an "addition-subtraction chain", in which metabolites are both synthesized and degraded within the pathway. Addition and addition-subtraction chains are concepts borrowed from the field of cryptography, whose relevance to our question will become apparent later. There is also a third, smaller category of cyclical pathways that cannot proceed unless a certain intermediate molecule is already present in the system. These pathways are autocatalytic cycles (Fig. 2B) that very much resemble autocatalytic cycles found in real biochemistry, such as the reverse TCA cycle [7], or the formose reaction [28]. Our results show that autocatalytic cycles can be simultaneously optimal for multiple tasks (Fig. 2B), suggesting that such types of structure may have a fundamental evolutionary advantage in a biological context. In addition to the recurrence of these topological categories among MBPs, we find that some specific structures are used repeatedly, often in a modular fashion (Fig. 2C). Specifically, many simple MBPs are used hierarchically as a toolkit for the construction of progressively more complex MBPs (data not shown), similar to what has been observed in real metabolic networks [29, 13, 30].

This modular architecture of recurring graph types provides a topological signature of optimally efficient pathways in our idealized chemistry. Since these pathways are chosen based on their minimal length, one may expect that a systematic analysis of all MBP lengths will display additional distinctive properties. Indeed, pathway lengths increase roughly logarithmically with the size of the input (or output) molecule (Fig. 3), with superimposed sharp jumps. For example, the task $a_9 \Rightarrow a_6$ can be performed in 2 steps, but the neighbor task $a_9 \Rightarrow a_7$ requires a minimum of 6 steps. Moreover, while most MBPs have only one or a few optimal realizations, selected instances display a peak in possible redundant solutions (Fig. S2), usually due to interconversions between molecules of similar size (e.g., $a_x \Rightarrow a_{x+1}$), or to the inherent complexity of a specific molecule (e.g., $a_7 \Rightarrow a_j$). These regular patterns suggest that it may be possible to reproduce the MBP length curves without having to actually compute the MBPs.

A similar search for patterns associated with minimal steps had been previously encountered in the mathematics of addition-subtraction chains, of high importance in cryptography [27]. These are integer sequences, beginning with 1, in which the *i*-th entry is either the sum or difference of any two previous entries in the sequence. These chains are often used in calculating large exponents of numbers [31]. For example, calculating $n^{128}$ can either be performed in 127 multiplications ($n \times n = n^2$, $n^2 \times n = n^3$,…, $n^{127} \times n = n^{128}$) or in a chain of 7 exponent multiplications ($n \times n = n^2$, $n^2 \times n^2 = n^4$, $n^4 \times n^4 = n^8$,…, $n^{64} \times n^{64} = n^{128}$). The latter can be further simplified by tracking the sums of the exponents in each calculation, which form an addition chain (1, 2, 4, 8, 16, 32, 64, 128). Shortest addition-subtraction chains are commonly used to calculate very large numbers in the fewest number of steps, thus speeding up computation time. These are often applied to methods in cryptography where the calculated exponents can be on the order of thousands to tens of thousands of bits [32].

The pathways explored in our model resemble optimal addition-subtraction chains. For example, the problem of obtaining $a_{128}$ from $a_1$ is formally equivalent to the addition chain example described above. However while typical addition-subtraction chains start with the number 1, in our MBPs we explore minimal paths starting from any molecule $a_i$ ($i \geq 1$). As described in detail in the Methods, we extended previous work on addition-subtraction chains [27, 33] to derive the following analytical estimate of the length of MBPs :

$$L(i, j) \sim \log_2 \frac{i}{\gcd(i, j)} + \log_2 \frac{j}{\gcd(i, j)} \tag{1}$$

where $L(i, j)$ is the number of reactions in the MBP with input $a_i$ and output $a_j$, and $\gcd(i, j)$ is the greatest common divisor of $i$ and $j$. As seen in Fig. 3, Eq. 1 reproduces the corresponding pathway lengths obtained by computing individual MBPs. This agreement implies that the number of reaction steps needed to construct an efficient metabolic pathway between two metabolites in our artificial chemistry can be roughly estimated from Eq. 1. The only feature that determines the pathway lengths is the complexity of the input and output molecules.

We can now ask whether similar minimal pathway length signatures are discernible in real metabolic networks. To cope with the gap in complexity between our model and real chemistry, we mapped real metabolic networks onto a single atom backbone [11, 12]. For example, the aldolase reaction, which cleaves fructose-1,6-bisphosphate ($C_6H_{14}O_{12}P_2$) into dihydroxyacetone phosphate ($C_3H_7O_6P$) and glyceraldehyde-3-phosphate ($C_3H_7O_6P$), can be mapped onto a carbon atom backbone, becoming simply $C_6 \leftrightarrow C_3 + C_3$ (see Methods). This reaction is now formally analogous to the $a_6 \leftrightarrow a_3 + a_3$ reaction in the idealized chemistry. Upon performing this mapping onto a carbon atom backbone, we ask whether the structure of real metabolic networks allows interconversions that use the minimal, logarithmic number of steps found for the artificial chemistry (Fig. 3 and Eq. 1). Specifically, we identified all shortest pathways between any two carbon compounds in *Escherichia coli*'s metabolic network. This was performed using two methods. The first was an explicit use of elementary flux modes as done in the artificial chemistry. As with the artificial chemistry method, this has the advantage of finding all of the shortest pathways that connect any two carbon compounds, but is limited in computational scope to a smaller network. Because of this limitation, we used the network of *E. coli*'s central carbon metabolism [34, 35], modified to remove cofactors and reactions that do not affect carbon transfer (see Methods). After finding all minimal elementary flux modes that connect every pair of carbon compounds in the network, we reduced those compounds to their carbon content alone, as described above. We determined, for each input compound, the length of the shortest elementary flux mode that reaches its closest molecule with $j$ carbons; then, for each value of $i$, we averaged these path lengths over all input molecules with $i$ carbons. The results show that the lengths of the *E. coli* elementary flux modes correlate with the lengths of the corresponding artificial chemistry MBPs and with the analytical predictions, though the actual *E. coli* values are overall larger than the artificial chemistry ones (Fig. 4). This last fact, as discussed later, may be due, for example, to energetic constraints, or to the higher complexity of real organic chemistry.

The second method is aimed at identifying all shortest pathways between any two carbon compounds in the whole genome-scale metabolic network of *Escherichia coli* [36], for which it is still infeasible to apply the elementary flux mode analysis. For this, we implemented a heuristic approach to analyze the set of shortest pathways between every pair of metabolites. We first determined, for each input compound, the minimal path length to reach its closest molecule with $j$ carbons; then, for each value of $i$, we averaged these path lengths over all input molecules with $i$ carbons. The results (Fig. 3) show that these *E. coli* minimal path lengths approximately follow the predicted logarithmic trend. For some curves (e.g. the one with $C_5$ as an input), the specific peaks and valleys of the predicted function are closely followed by the *E. coli* network. While this does not prove that MBPs are indeed used in real metabolic networks, it suggests that the logarithmic strategy of MBPs is embedded in their architecture. However, because this second method focuses on shortest paths across a metabolite-to-metabolite network rather than on flow-conserving MBPs, the predicted values are likely an underestimate of the number of reactions necessary to construct one metabolite from another in a mass-conserved manner.

So far, we have analyzed the properties of individual MBPs in our idealized chemistry, as well as analogous minimal length pathways in *E. coli* metabolism. However, some fundamental aspects of the architecture of metabolism may be visible only at the ecosystem-level, namely by collectively analyzing the metabolic network obtained as the union of the metabolic maps of known individual species (sometimes called the "meta-metabolome"). For example, previous work using an algorithm of network expansion applied to this meta-metabolome has identified potential signatures of major evolutionary events [37], including the metabolic transition that took place upon the great oxidation event, about two billion years ago [4].

Here, we build a meta-metabolome for our idealized chemistry by considering the collection of MBPs. One could imagine that each task $a_i \Rightarrow a_j$ corresponds to a different organism, which has filled a specific metabolic niche (availability of $a_i$), and found an optimal solution (the MBP) for its main metabolic task (produce $a_j$). The question we ask next is whether, in this ecosystem of MBPs, all metabolites and reactions are used in roughly the same number of pathways, or if specific metabolites or reactions seem to be essential for many optimal tasks, hence representing "universal tools".

For this analysis we used the set of MBPs calculated on the $R_{19}$ network using the MILP method. One first result of this analysis is that every metabolite of an even length is used in many more MBP reactions than their odd length neighbors, compared to the underlying chemistry (Fig. 5B). Thus even-length metabolites are more important in that they can be used for more tasks. A possible explanation for this enhanced importance comes from the logarithmic nature of the MBP path lengths. For example, producing $a_8$ from $a_1$ requires only three doubling reactions ($a_1 + a_1 \rightarrow a_2$, $a_2 + a_2 \rightarrow a_4$, and $a_4 + a_4 \rightarrow a_8$). In addition, this same pathway, with one additional reaction, can also be used to optimally produce $a_9$ and $a_{10}$ (see Table S2), overall increasing the number of pathways in which each of those even-length intermediates is used. Indeed, because similar logarithmic pathways can be used as a backbone connecting distant inputs and outputs, we expect metabolites of an even length to appear more often. Similarly, one can address the relevance of each possible reaction across different MBPs. The existence of ubiquitous reactions is visible in Fig. 2A and Fig. S1, and can be more systematically assessed by plotting a usage distribution (Fig. S2). The most abundant reactions – the "universal tools" in this model chemistry – are the ones that ligate two identical molecules (e.g. $a_2 + a_2 \leftrightarrow a_4$, see Tables, 1, S2, and S3). Strikingly, the distribution of reaction utilization follows a long-tailed distribution (Fig. 5A), whose fit to a power law gives an exponent of approximately -1.1 ($R^2 = 0.99$). This value is close to our theoretically predicted value of -1 (See Methods).

As in the case of the artificial chemistry network, we can now search for patterns of metabolites and reactions usage in the collective set of all metabolic reactions known in living systems, obtained from the KEGG database [15]. The presence of such signatures would suggest a long-term selective advantage of molecules and reactions that are useful for multiple tasks across different organisms and environments. By counting how many times each possible carbon backbone reaction is used across this biosphere-level metabolism we obtained a broad distribution, and a fit to a power-law gives the exponent of -0.89, comparable with the analytically predicted value, and with that in the artificial chemistry model (Fig. 5A and Fig. S4). Most surprisingly, we found that several reactions that are top ranking in their count across MBPs in the artificial network, are also at the top of the list in the KEGG-derived reactions (Spearman correlation p-value<$10^{-6}$; see also Table 1, and Supplementary Spreadsheet). This suggests that the $R_N$ network model, despite its simplified chemical rules, captures some fundamental features of the role of the carbon reaction backbone of real metabolic networks.

In addition to a preference for specific reactions, we can ask whether the spectrum of metabolite usage across the whole KEGG metabolism reflects the possible optimality criteria encountered in the model (Fig. 5C). The metabolite usage in the hydrogen backbone network (see Methods) is similar to that in the artificial chemistry: each even-length hydrogen metabolite is used more often than its odd-length neighbors (Fig. S4A). For the carbon backbone distribution, we see a similar descending periodic behavior, but with a periodicity of approximately 5 (Fig. 5C and Fig. S5B). Hence, molecules containing carbons in a number that is a multiple of 5 are used more abundantly than other molecules across different metabolic reactions. One possible explanation for this $C_5$ periodicity is the profuse usage of adenine and nicotinamide adenine dinuculeotide compounds as energy carriers and redox balance molecules, although the removal of such compounds has little effect on the observed periodicity (Fig. S6). Hence, the prominent usage of compounds with specific numbers of carbons might reflect global network optimization principles for the efficiency of multiple pathways, as observed in the artificial chemistry model. The periodicity of 5 that we observe, together with the evidence displayed in Fig. 3C, may suggest that the evolutionary optimization of metabolism has been partially taking place around building blocks of five carbons, compatible with previous observations of prebiotic abundance of terpenoids [38] and pentoses [39]. It is also interesting to note that an unexplained periodicity of two had been previously observed in the distribution of the number of carbons among known organic compounds [40-42]. While our analysis is based on the distribution of usage of carbon compounds in different reactions, rather than the total count of molecules, future analyses may investigate possible connections between these trends.

**Discussion**

We have explored the potential existence of general principles underlying the evolution of metabolic network architecture. Specifically, we studied the properties of pathways (the MBPs) that perform elementary metabolic tasks with maximal yield and minimal length in an idealized chemistry. Using the results from the model chemistry, we asked whether similar signatures of optimally efficient organization could be found in real metabolic networks.

In computing possible MBPs, we have focused mostly on identifying modular features, on predicting their lengths, and on the statistics of usage of metabolites and reactions. In the future, it may be interesting to characterize the full spectrum of degenerate MBPs for large artificial chemistries. This would allow to assess, for example, the density of specific topologies (such as the autocatalytic cycles), or the dependence of degeneracy on the numerical properties of input/output pairs. One of our algorithms (the elementary modes one) can find a large number of degenerate solutions, including autocatalytic cycles. This algorithm is currently not scalable, because of the difficulty of computing elementary flux modes, especially in the highly connected artificial chemistry network we have used, though very recent improvements in elementary flux mode calculations [44] might be useful for this enumeration. In addition, while intuitively this approach seems to capture the full degeneracy of MBPs, this still remains to be formally proven. Another intriguing possibility might be to modify our flux balance MILP approach to identifying degenerate solutions by employing integer cuts, as described in [45]. Alternative avenues for optimization using Linear Programming rather than MILP could be in principle devised to reduce the complexity of calculations. For example minimizing the sum of absolute values of fluxes allows for rapid calculation of pathways up to the $R_{100}$ network, though in this case the ensuing pathways are not of minimal length (see Fig. S7). In any case, for the purpose of the current work, we verified that degeneracy does not affect the statistics of usage of different reactions (Fig S8).

Among the recurrent MBP topologies identified, we encountered numerous autocatalytic cycles. The properties of autocatalytic cycles have been studied previously [9, 46], and their self-replication potential has been theorized to be important in the early evolution of carbon fixation [8, 7]. Autocatalytic cycles have also been shown to be kinetically stable, even in the absence of regulatory control [47]. We found that some autocatalytic pathways (e.g. the pathway from $a_7$ to $a_8$) are simultaneously optimal for multiple metabolic tasks. In this specific case, the MBP $a_7 \Rightarrow a_8$ is also an MBP for the production of each intermediate in the cycle (Fig. 2B). This special property of autocatalytic cycles in our artificial chemistry may have a parallel in real metabolism. For example, many metabolites in the TCA cycle (which is autocatalytic when run in reverse [7]) are precursors for fundamental anabolic processes [48, 49]. Similar properties can be observed in the fundamental autocatalytic cycle known as the formose reaction [28].

Along with the structural details of MBPs, we also used analytical methods to estimate the length of MBPs as a function of the length of input and output molecules. This estimate closely matched the lengths of the artificial chemistry pathways computed with numerical algorithms. These calculations establish a new link between two apparently unrelated disciplines, namely the mathematics of addition-subtraction chains and biochemistry. It will be interesting to explore in the future whether extensions to more realistic artificial chemistries can be formalized in a similar fashion. Conversely, the MBP length estimate obtained for biochemical pathways may suggest new avenues in applied mathematics.

To determine whether predictions of minimal MBP length in our idealized chemistry could have implications in real biochemistry, we searched for pathways of minimal length between compounds with different counts of carbon atoms in the *E. coli* metabolic network. The complexity of real chemistry relative to our idealized system made this comparison difficult to interpret. MBPs in *E. coli* were found to be composed of many more reactions than in the artificial chemistry. This might be due to the additional requirement of energy gradients (e.g. the phosphorylation steps), to the complex interdependence of multiple elements, and to the properties of stability of intermediates. Previous work had addressed the question of optimality in specific metabolic pathways. For example, Melendez-Hevia and Torres [43], used optimality criteria to infer that some metabolic pathways, most notably the pentose-phosphate pathway, can be traversed using the fewest number of reactions. Heinrich and Schuster [50], conversely, describe the identification of a series of phosphorylation/dephosphorylation and ATP consumption/production steps that maximize the flux of ATP production in central carbon metabolism pathways. In contrast, our search for MBPs in the artificial chemistry model corresponds to a systematic search for maximal yield, minimal length pathways for all possible input-output relationships (something not yet feasible for real metabolic pathways). This allows us to infer analytical relationships and potential principles that may hold (perhaps in an approximate way) for virtually any evolved metabolism. In real systems, various compounds are often produced as waste byproducts and simply excreted into the environment, leading to suboptimal yields of target metabolites. For example, one could go in one step from a $C_6$ compound to $C_5$, with a yield of 5/6 ~ 83% and an additional $C_1$ byproduct; yet, obtaining maximal yield in this transformation, would cost at least 3 more reactions. One could argue that the combination of all of these criteria, including the maximization of ATP production and optimization of enzymatic catalysis may have played a key role in the evolution of modern metabolism, leading to compromise solutions. Exploring pathways that produce multiple compounds from multiple inputs with the addition of thermodynamic constraints might constitute an interesting model extension for further investigation.

We found that the statistical properties of the usage of reactions across MBPs recapitulate the statistics of reaction usage in the union of all known metabolic pathways (represented by the KEGG metabolic database). Both across the set of all MBPs for the idealized chemistry, and in the KEGG metabolic map, we observed that a few reactions are used far more often than many of the others in the set. Another way of determining the importance of individual reactions in the context of the global functionalities of a meta-metabolome would be to perform perturbation experiments. We implemented such an experiment in our idealized chemistry, by progressively removing reactions and checking how many metabolites can still be produced. Depending on whether reactions are removed in random order or in the order determined by their usage across MBPs, the outcome is quite different (Fig. S9). This analysis sheds light on the importance of reactions in terms of the capacity to produce a certain output.

Determining to what extent real metabolic networks obey optimality principles like the ones described here will take additional effort. Even if an underlying arithmetic simplicity governs idealized optimal pathways, deviations from ideal behavior should be expected. For example, parallel selection pressures for energy production and biochemical stability would likely sacrifice pathway minimality. However, guiding principles as the ones we are proposing could serve as reference points for future research, including circumstances in which metabolism can be different from what we are used to. Using synthetic biology techniques, for example, it might be possible to redesign metabolic pathways so as to approach predicted ideal efficiencies and minimal enzyme cost [51, 52]. From a totally different perspective, in the field of astrobiology, having a prediction of possible signatures of an evolved metabolism might help select, among the molecular spectra of extrasolar planets, those possibly indicative of biogenic processes [53-55].

**Materials and Methods**

*1. Artificial Chemistry Model*

We define an artificial chemistry inspired by previous string-based artificial chemistries (see also main text and Fig. 1). One may think of molecules in this artificial chemistry as polymers (up to a given length $N$) of a monomeric unit $a$. Since no specific assumption is made in the model about the nature of these molecules, they could equally represent aggregates or branched polymers of different sizes, as well as molecules with different counts of a specific atom. A network $R_N = \{M_N, C_N\}$ is defined by the set of $N$ molecules $M_N = \{a_i \mid \forall\ i = 1,\ldots,N\}$ and the set of all possible uni-bi ligation/lysis reactions between them, $C_N = \{a_i + a_j \leftrightarrow a_k \mid \forall\ i, j,$ and $k$, such that $i \leq j, i + j = k,$ and $k = 2,\ldots,N\}$.

*2. Flux Balance Analysis*

Flux Balance Analysis (FBA) is a steady state constraint-based approach to study the flow of mass through metabolic networks [26, 56, 57]. Briefly, FBA represents the metabolic network of interest as an $n \times m$ stoichiometric matrix $S$, whose element $S_{ij}$ indicates the number of molecules of metabolite $i$ ($i=1,\ldots,m$) that participate in reaction $j$ ($j=1,\ldots,n$) (with a positive sign if the metabolite is produced, negative if it is consumed). Each reaction can be associated with a rate, or flux, $v_j$. Under the assumption of a steady state the following set of mass conservation constraints on the fluxes is generated:

$$\sum_{j=1}^{n} S_{ij} v_j = 0 \qquad i = 1, 2, \ldots, m \qquad (1)$$

Additional constraints (such as availability of nutrients, experimentally observed irreversibility, maximal or minimal rates, etc.) can be imposed on the fluxes as inequalities of the form

$$\alpha_j \leq v_j \leq \beta_j \qquad (2)$$

where $\alpha_j$ is the minimal allowed rate of a reaction and $\beta_j$ is its maximal rate. Taken together, the above constraints define a convex polyhedron (the "feasible space") in the $n$-dimensional space of fluxes. Linear programming (LP) can be used to identify, within the feasible space, flux vectors that maximize or minimize a given linear objective function. In microbial systems it has been often hypothesized that a biologically meaningful objective is the maximization of the flux through the reaction that represents cellular growth, or biomass production [2, 58]. Hence, LP applied to FBA provides a prediction of all metabolic fluxes in a cell. FBA can be applied at genome scale, and corresponding stoichiometric models are available for a number of organisms. FBA predictions have been experimentally validated most thoroughly in *Saccharomyces cerevisiae* SC288 [59] and *E. coli* K-12 [36].

*3. Minimal Balanced Pathway discovery algorithms.*

Minimal Balanced Pathways (MBPs) are defined as sets of reactions in the $R_N$ network that can optimally perform a given metabolic task. A task is defined as the production of a specific end product (e.g., $a_j$, with output flux $v_{out}$) from a single available nutrient (e.g., $a_i$, with input flux $v_{in}$). A pathway between two molecules is a MBP if (i) it satisfies a steady state solution, analogous to Eq. 1; (ii) it produces the final product with maximal yield, i.e., $v_{out}=v_{in} \cdot j/i$; and (iii) it contains the smallest possible number of reaction steps. The MBP between $a_i$ and $a_j$ will be indicated as $a_i \Rightarrow a_j$.

We have developed three different algorithms for computing MBPs, as described below:

a. Flux Balance Analysis/Mixed Integer LP algorithm

We use a modified FBA approach to formulate the MBP problem in a constrained optimization framework. Specifically, we impose the same constraints used in an FBA problem, and further require that the maximal yield condition $v_{out}=v_{in} \cdot j/i$ be satisfied. We then search for a solution that minimizes the number of active (nonzero) fluxes. Towards this goal, we use a modification of the LP problem described above to introduce binary variables ($b_j$) that represent flux activity: $b_j=0$ if $v_j=0$, and $b_j=1$ otherwise. To identify a minimal path, we can then search for the set of fluxes that minimize $\Sigma_i b_i$. Because of the nature of the variables involved – the fluxes are continuous, and the number of active fluxes is an integer – this problem must be solved using a mixed integer linear programming (MILP) algorithm. Our MILP problem for the optimal MBP between $a_n$ and $a_m$ can be formulated as follows:

Minimize $\sum_{j=1}^{N} b_j$

Subject to:

$$\sum_{j=1}^{n} S_{ij} v_j = 0 \quad (5)$$

$$\alpha_j b_j \leq v_j \leq \beta_j b_j \qquad \text{for } j = 1, 2, \ldots, n$$

$$v_{out} = \frac{m}{n} v_{in}$$

$$b_j \in \{0,1\} \qquad \text{for } j = 1, 2, \ldots, n$$

The optimal solution for this problem will give the flux distribution $v$ that uses the fewest nonzero values to maximize the objective. In our MBP computations, the only flux constraints used were those that limit the uptake of the single nutrient to an arbitrary value of 10 mmol/grDM·h, and the production of the target metabolite to the known maximal yield $v_{out}/v_{in}=j/i$.

b. Elementary Flux Modes algorithm

Given a metabolic network defined by a stoichiometric matrix $S$ (as described in the above FBA section), a vector of fluxes $v$ is said to correspond to an Elementary Flux Mode (EFM) if it satisfies the following three conditions [21].

1. It satisfies the steady state condition ($Sv = 0$).
2. It must be feasible within the conditions of the model: if there are known boundaries for the fluxes, then $v$ must fall within them.
3. It must be non-decomposable. There are no two smaller EFMs that can be linearly combined to form the one in question.

Because of these constraints, those EFMs that use the minimal number of reactions satisfy the requirements for being an MBP. We used the METATOOL software package [60] to find all EFMs in the $R_{10}$ network, and then identified all of those EFMs that are also MBPs.

c. Iterative additive algorithm

We designed and implemented an algorithm to produce most MBPs *de novo*, without relying on prior steady state stoichiometric modeling methods. The algorithm works in an iterative manner, producing longer pathways from shorter ones. For example, we can start from two trivial MBPs: $a_1 \Rightarrow a_1$ (which requires no reactions), and $a_1 \Rightarrow a_2$ (requiring one trivial reaction, $a_1 + a_1 \rightarrow a_2$). To compute $a_1 \Rightarrow a_3$, we identify all the ways in which we can decompose 3 into two smaller addends (in this case, only one: 3=2+1). Next we combine together the previously computed MBPs that progress from $a_1$ to each of these two addends, giving a new putative MBP for the desired new task ($a_1 + a_1 \rightarrow a_2$, and $a_1 + a_2 \rightarrow a_3$). This procedure can be then iterated to give a prediction of MBP $a_i \Rightarrow a_j$ ($i, j \leq N$).

       This algorithm is fast and efficient compared to the previous methods, allowing us to apply it to even the $R_{100}$ network. However, it has two main drawbacks. First, it will miss pathways that "overshoot" the target value then subtract down to it. Second, it may miss MBPs that are not built modularly from smaller ones. From a comparison of the MBPs predicted by the different algorithms, one can see that the approximations introduced in this algorithm cause 18 out 361 MBPs (5%) in $R_{19}$ to overestimate pathway length by one reaction. Also, this algorithm correctly identifies 204 of the 384 degenerate MBPs that the EFM algorithm finds in $R_{10}$. The reaction usage using this method is highly correlated with that of the MILP method applied to $R_{19}$ (Pearson correlation 0.96, p-val $10^{-51}$), and the EFM method applied to $R_{10}$ (Pearson correlation 0.98, p-val $2 \cdot 10^{-17}$).

*4. KEGG reaction reduction*

Data used for the comparison between the $R_N$ metabolic network and real metabolism was gathered from the KEGG LIGAND database (July 26, 2009 release)[15]. This database was parsed to convert its compounds and reactions into a single-atom form, as described in the text. Compounds that carried any uncertainty in their atomic makeup, including non-specific side-chains or variable chain length were removed from the current analysis. We also removed from the analysis reactions with no associated formula, as well as reactions involving non-specific molecules (such as generic glycans and non-specific nucleotide or peptide chains). Finally, a number of reactions were found to leave the atomic composition of the compounds essentially unchanged on either side of the reaction (e.g., $C_3 \leftrightarrow C_3$). These reactions were ignored as well, without consequences on the results (data not shown).

*5. Metabolite and reaction usage*

We counted how often each metabolite and reaction was used in the artificial chemistry pathways as well as in the KEGG-derived single-atom networks. In the model pathways, reaction usage was calculated by counting how many times each reaction was used across all pathways. Metabolite usage was similarly calculated by counting the occurrence of reactions in which each metabolite participates. For example, in the pathways that convert $a_9$ to $a_{10}$ in Fig. 2C, $a_9$ participates in only one reaction, but $a_{10}$ participates in two.

In the KEGG-derived networks, a similar counting scheme was used. The reaction usage was calculated by counting how many times each reduced reaction appears, and the metabolite usage was calculated by counting how many times each metabolite appears across all reactions.

*6. Shortest paths in* Escherichia coli

a. Elementary flux modes

The first method used EFMs to find all shortest pathways in the central carbon metabolism of *Escherichia coli* [34, 35]. Because we are interested in the pathways that alter the carbon content of different molecular species, we removed those common cofactors that do not alter the carbon content of other metabolites (ATP, ADP, AMP, NADH, NADPH, NAD+, NADP+, H+, and $PO_4$). Also, we effectively ignored reactions involving transport and exchange. We then used the EFM method described in Methods 3c above to find the number of reactions in each of the MBPs for this reduced network.

For each input compound, we listed the lengths of the MBPs for all output compounds containing a number $j$ of carbons. For each $j$, we select among these paths the shortest one, giving an estimate of the shortest path between any individual compound and the closest $j$-carbon compound. Finally, for each value of $i$, we averaged these path lengths over all input molecules with $i$ carbons. The end result is a matrix that provides the average of the shortest paths from any $i$-carbon compound to its nearest $j$-carbon compound.

b. All-pairs shortest paths with Johnson's algorithm
The larger, genome-scale metabolic network of *E. coli* [36] has 761 metabolites and 1075 reactions, and is currently infeasible to study with the EFM method described above, limiting us to a graph theoretical approach. Because there are both metabolites and reactions, metabolic networks are inherently bipartite: metabolites connect to each other only through reactions. Pathway lengths were computed by transforming the metabolic reaction network stoichiometric matrix into an adjacency matrix where metabolites and reactions are all represented by the same type of node.

As described above, we are only interested in the connections between carbon compounds, so we removed any non-carbonaceous metabolites (water, phosphate, ammonia, etc.). Also, we removed the following cofactors that are used in many reactions, but do not participate in the transformation of carbons: ATP, ADP, AMP, NAD+, NADH, NADP+, NADPH, coenzyme A, acetyl-CoA, and the acyl carrier protein.

Next, we used Johnson's all-pairs shortest paths algorithm (available as a Matlab function) to find shortest pathways between any two carbon compounds in *E. coli*'s metabolic network. This set of pathways was collated as in the previous section, to produce a matrix of the average of the shortest paths from any $i$-carbon compound to its nearest $j$-carbon compound.

*7. Analytical estimate of MBP lengths in analogy with addition-subtraction chains*
We developed an analytical approximation for the expected numbers of reactions to be found in any MBP $a_i \Rightarrow a_j$. We begin with a simplified version of the artificial chemistry model in which only irreversible addition reactions of the form

$$a_p + a_q \to a_{(p+q)} \qquad (1)$$

are allowed. Under these restrictions, we first ask what is the smallest number of reactions necessary to produce any $a_j$ from $a_1$. We shall denote by $l(j)$ the smallest possible number of such reactions (we count the use of each reaction (1) once). This problem is equivalent to the problem of addition chains [27], in which one attempts to compute a positive integer by generating a sequence of integers such that each term in the sequence is the sum of two previous terms. Addition chains have been studied extensively, mainly because of their applications in computer science and cryptography [27]. For addition chains, $l(j)$ grows logarithmically with $j$:

$$l(j) \propto \log_2(j) \qquad (2)$$

Our artificial chemistry represents a generalization, in which a metabolite of any length $i$ can be used to produce an output metabolite of any length $j$. If we still assume that only addition reactions are possible (i.e. molecules cannot be broken down), a chain from $a_i$ to $a_j$ will exist only when $i$ is a divisor of $j$. The problem can then be reduced to the case with $a_1$ input and $a_{j/i}$ output. Therefore, in the irreversible case, we can assume that inputs consist of monomers without loss of generality. Let $L(j)$ be the length of the shortest reaction chain in this case. Because not every reactant exists when dividing by the input length $i$, we have the obvious inequality

$$l(j) \leq L(j) \qquad (3)$$

Sometimes the shortest chain can be found easily. For instance, $\{2^0, 2^1, ..., 2^k\}$ is obviously the shortest chain from 1 to $j = 2^k$ whose length is $k + 1$. This suggests the general lower bound on the shortest length $L(j)$ of the addition chain:

$$L(j) \geq \lceil \log_2(j) \rceil + 1 \qquad (4)$$

where $\lceil x \rceil$ represents the ceiling of $x$, or the smallest integer not less than $x$. Likewise, as seen below, $\lfloor x \rfloor$ represents the floor of $x$, or the largest integer not greater than $x$ (for example, $\lceil 3.14 \rceil = 4$, and $\lfloor 3.14 \rfloor = 3$). The longest minimal addition chain arises when the output length is $j = 2^m-1$. From this fact, we have the upper bound [27]

$$L(j) \leq \lfloor \log_2(j) \rfloor + \upsilon(j) \qquad (5)$$

where $\upsilon(j)$ is the number of 1s in the binary representation of $j$. Since $\upsilon(j) \leq \lfloor \log_2(j) \rfloor + 1$, the bound in equation (5) implies a simpler (but weaker) upper bound

$$L(j) \leq 2\lfloor \log_2(j) \rfloor + 1 \tag{6}$$

The above bounds give precise values in some cases and act as bounds in others. For instance, $L(16) = 5$, and $L(17) = L(18) = 6$ for both the lower and upper bounds, while $L(31) = 8$ is between the lower bound and the upper bound (5 and 9, respectively).

There are various conjectures regarding $L(j)$; one of the most famous [33] asserts that computing $L(j)$ is NP-hard. Nonetheless, the computation of $L(j)$ has been pushed up to $n \leq 2^{25}$. Two other conjectures [31] predict the general lower bound

$$L(j) \geq \lfloor \log_2(j) \rfloor + \log_2 \upsilon(j) + 1 \tag{7}$$

and the upper bound

$$L(2^k - 1) \leq k + L(k) - 1 \tag{8}$$

While algorithms for generating the shortest addition chains are discussed by Thurber [31], these all hold for the specific case of pure addition where the input is always $a_1$.

We are interested in the general case involving both addition and subtraction, and specifically the lengths $l(i, j)$ of the shortest reaction chains (MBPs) with $a_i$ input and $a_j$ output. Addition-subtraction chains have also been studied previously as an expansion of addition chains, although these correspond to MBPs with only $a_1$ as an input. Sometimes, in these cases, $l(j)$ is readily computable, e.g.

$$l(2^k - 1) = k + 2 \qquad \text{for } k \geq 3 \tag{9}$$

while $L(2^k - 1)$ remains unknown for sufficiently large $k$. Both lengths can also be equal, i.e. $l(j) = L(j)$. For example,

$$l(2^k) = L(2^k) + k = 1 \tag{10}$$
$$l(2^k + 1) = L(2^k + 1) = k + 2 \tag{11}$$

Note also an inequality:

$$l(j) \geq \lceil \log_2(j) \rceil + 1 \tag{12}$$

All of these features explain the growth law in equation (2).

The quantity $l(i, j)$ has a rich behavior, e.g., there is only a trivial lower bound since $l(j, j) = 1$. To ignore this non-interesting effect, let us divide $i$ and $j$ by their greatest common divisor as it never affects the length of the MBP:

$$l(i,j) = l\left(\frac{i}{d}, \frac{j}{d}\right), \qquad d = \gcd(i,j) \tag{13}$$

then we can use an obvious inequality

$$l(i,j) \leq l(i,1) + l(1,j) = l(i) + l(j) \tag{14}$$

Recalling (2) we finally arrive at an approximation for the number of reactions in an MBP that uses $a_i$ to produce $a_j$:

$$l(i,j) \sim \log_2\left(\frac{i}{d}\right) + \log_2\left(\frac{j}{d}\right) \tag{15}$$

The approximation in (15) can also be used to estimate the rank distribution of reaction usage. Consider all possible MBPs producing $a_j$ from $a_i$. For each $(i, j)$ pair, take an MBP and mark all reactions. Let the reaction in (1) occur $E_{pq}$ times: that is, there are $E_{pq}$ MBPs that use (1). We now divide $E_{pq}$ by the total number of MBPs and call $e_{pq} = N^{-2} E_{pq}$ the reaction frequency. It is better to order reactions not according to $(p, q)$ but to their ranking $j$, so that the reaction of rank $j = 1$ is the most frequent, that of rank $j = 2$ is the second in frequency, etc. This gives $e_j$. How does $e_j$ decrease with rank? To infer the answer we note that

$$\sum_{j=1}^{N^2} e_j = \langle l \rangle \tag{16}$$

From (15) it is clear that the average length $\langle l \rangle$ of the shortest reaction chain scales as log $N$. This is consistent with (16) if and only if we have $r_j \sim j^{-1}$. Thus we predict the power-law decay

$$r_j \sim j^{-1} \text{ when } j \gg 1 \tag{17}$$

**Table 1**

Of the top 10 most used reactions in the $R_{19}$ network and carbon-only KEGG network, there are five equivalent reactions that appear in both. Recognizing that 74 of the 90 possible reactions from the $R_{19}$ set are found in the 631 carbon-only reactions from KEGG, we can use the Spearman rank-correlation to find that this has a correlation value of 0.54 with p-value $8 \cdot 10^{-7}$.

|    | Model reaction | KEGG carbon reaction |
|----|----------------|----------------------|
| 1  | $a_2 + a_2 \leftrightarrow a_4$ | $C_6 + C_6 \leftrightarrow C_{12}$ |
| 2  | $a_1 + a_1 \leftrightarrow a_2$ | $C_1 + C_5 \leftrightarrow C_6$ |
| 3  | $a_4 + a_4 \leftrightarrow a_8$ | $C_1 + C_3 \leftrightarrow C_4$ |
| 4  | **$a_3 + a_3 \leftrightarrow a_6$** | **$C_1 + C_4 \leftrightarrow C_5$** |
| 5  | $a_2 + a_4 \leftrightarrow a_6$ | **$C_5 + C_5 \leftrightarrow C_{10}$** |
| 6  | **$a_6 + a_6 \leftrightarrow a_{12}$** | $C_1 + C_7 \leftrightarrow C_8$ |
| 7  | **$a_1 + a_2 \leftrightarrow a_3$** | $C_1 + C_8 \leftrightarrow C_9$ |
| 8  | $a_8 + a_8 \leftrightarrow a_{16}$ | **$C_3 + C_3 \leftrightarrow C_6$** |
| 9  | **$a_5 + a_5 \leftrightarrow a_{10}$** | $C_4 + C_5 \leftrightarrow C_9$ |
| 10 | **$a_1 + a_4 \leftrightarrow a_5$** | **$C_1 + C_2 \leftrightarrow C_3$** |

**FIGURE LEGENDS**

**Figure 1**

Representation of the $R_4$ network. (**A**) This schematic of the $R_4$ artificial chemistry network is composed of metabolite "strings" of $a$ up to a maximum length of four, and all allowed reactions between them.

(**B**) The reaction list for the $R_4$ network. There are four reactions that represent the exchange of mass with the environment ($r_1$-$r_4$) – one for each metabolite – and four reactions between the metabolites ($r_5$-$r_8$).

**Figure 2**

Emergent complexity and modularity in artificial network topology. (**A**) This set of example MBPs displays the emergent modularity of structure and function, as well as the multiple usage of different reactions. Each row denotes the input metabolite used, and each column the output metabolite. The reaction marked in red interconverts $a_2 + a_2 \leftrightarrow a_4$ and is the most used reaction. MBPs on a yellow background are autocatalytic cycles. For a larger image with more examples, and a more detailed view of the properties observed in these networks, see Figure S1. (**B**) A single autocatalytic loop is used as a modular backbone for several MBPs: the cycle that constructs $a_8$ from $a_7$ is also used to produce $a_1$, $a_2$, and $a_4$. (**C**) Four examples of the MBP that produces $a_{10}$ from $a_9$. Each breaks down $a_{10}$ into $a_1$ in different but equally optimal ways. Each of these sub-pathways (gray metabolites) is an MBP in itself, showing the modularity of use of each of these metabolic tools.

**Figure 3**

Logarithmic growth is observed among pathways from both the $R_{19}$ and *E. coli* metabolic networks. Each plot has a single starting value *i* corresponding to either $a_i$ (for the model) or $C_i$ (for *E. coli*). The red lines show the number of reactions in each MBP with different output metabolite size, and the black lines show the average number of reactions used to reach the nearest metabolite of increasing carbon number in *E. coli*. The predicted number of reactions from Equation 1 is also shown for each plot in blue. (**A**) $C_1/a_1$ input. (**B**) $C_2/a_2$ input. (**C**) $C_5/a_5$ input.

**Figure 4**

Lengths of MBPs in the $R_{19}$ network compared to minimal elementary modes in the *E. coli* central carbon metabolic network. Each plot has a single starting value *i* corresponding to either $a_i$ for the model or $C_i$ for *E. coli*. The red lines show the number of reactions in each MBP with different output metabolite size The black lines show the average number of reactions used to reach the nearest metabolite of increasing carbon number in *E. coli* central carbon metabolism. The predicted number of reactions from Equation 1 is also shown for each plot in blue. (**A**) $C_2/a_2$ input. Correlation coefficient = 0.92, p-val = 0.003. (**B**) $C_3/a_3$ input. Correlation coefficient = 0.94, p-val = 0.001. (**C**) $C_5/a_5$ input. Correlation coefficient = 0.78, p-val = 0.04.

**Figure 5**

Metabolite and reaction usage frequencies. (**A**) The frequency of usage of reactions in the artificial chemistry model and in the KEGG-derived carbon reaction set. MBPs for the $R_{19}$ network were calculated using the MILP method while the others ($R_{30}$ through $R_{100}$) were estimated using the iterative algorithm. We calculated the reaction usage by counting the number of MBPs that use each reaction. These were then ranked in descending order, yielding curves that follow a power law with an average exponent of -1.14 (+/- 0.03) ($R^2$=0.99). The reaction usage in the KEGG-derived carbon dataset was calculated by counting the number of times each equivalent reaction appears, and follows a power law tail distribution, with exponent -0.89. The curve predicted by the analytical model, with exponent -1, is shown as a solid line. (**B**) The usage frequency of each metabolite among all MBPs in the $R_{19}$ model. This also shows the frequency of use of each metabolite in the $R_{19}$ network itself, and in a randomly chosen set of reactions (control). In the inset, the metabolite usage was sorted by rank and plotted on a semilog axis. (**C**) The usage frequency of each metabolite among all reactions in the KEGG carbon reaction set. The inset shows the usage sorted by rank on a semi-log scale.

# Figure 1

**A**

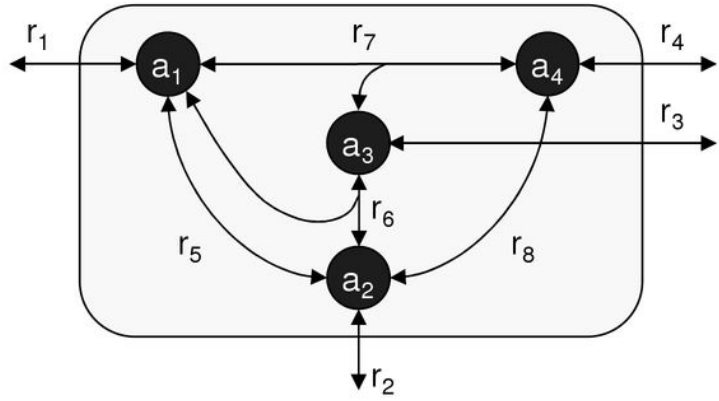

**B**

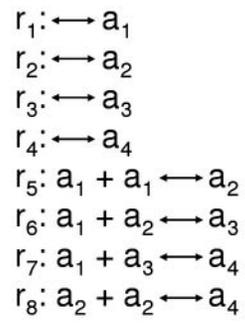

$r_1: \longleftrightarrow a_1$
$r_2: \longleftrightarrow a_2$
$r_3: \longleftrightarrow a_3$
$r_4: \longleftrightarrow a_4$
$r_5: a_1 + a_1 \longleftrightarrow a_2$
$r_6: a_1 + a_2 \longleftrightarrow a_3$
$r_7: a_1 + a_3 \longleftrightarrow a_4$
$r_8: a_2 + a_2 \longleftrightarrow a_4$

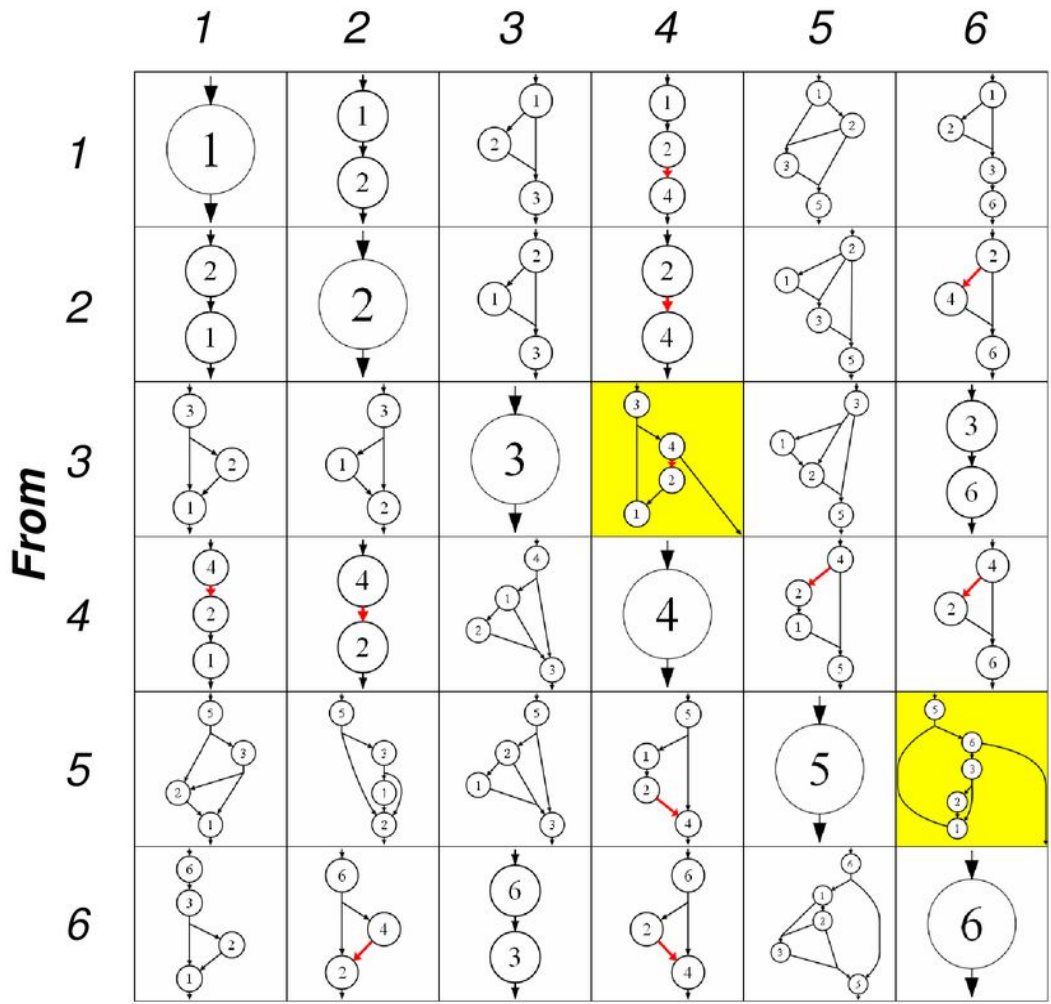

**A**

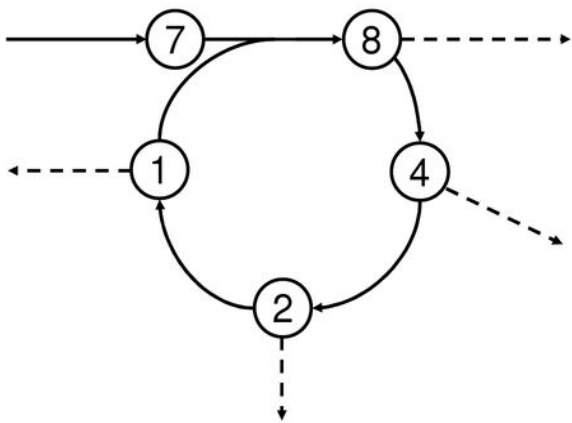

**B**

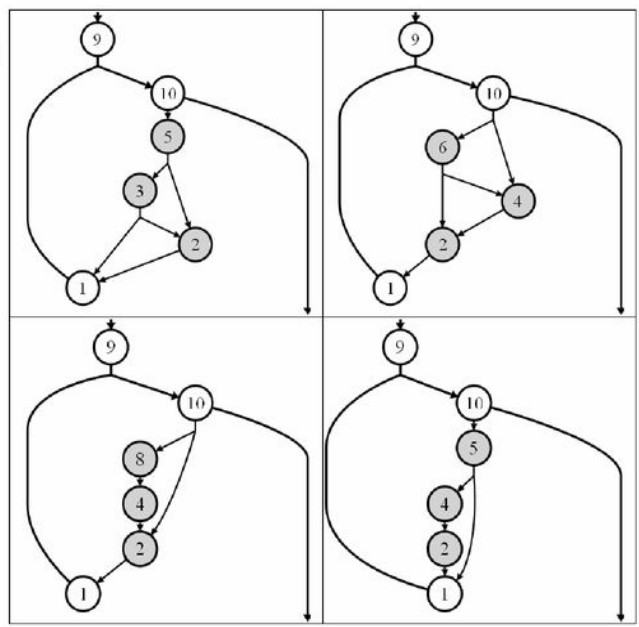

**C**

# Figure 3

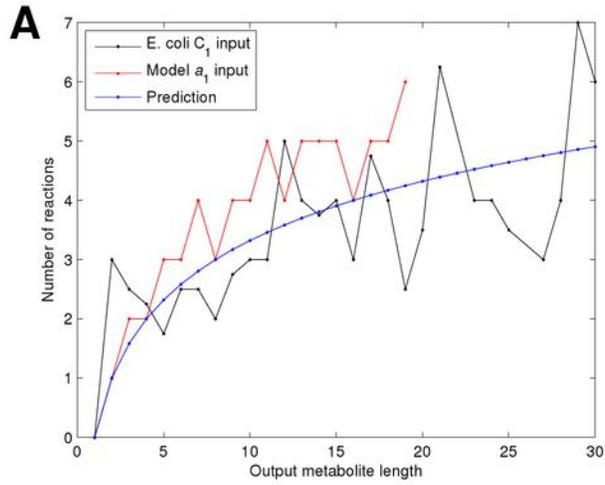

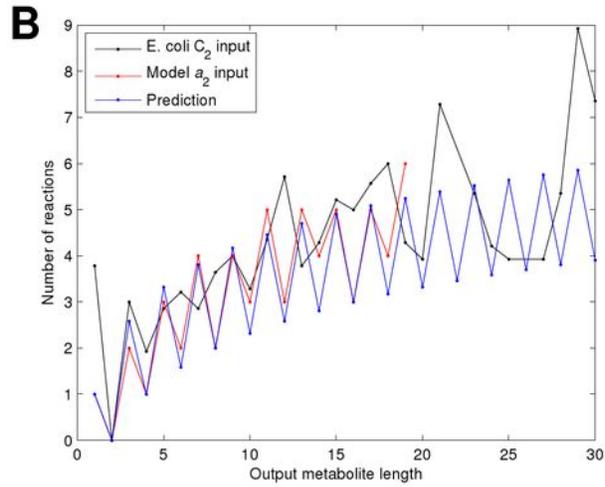

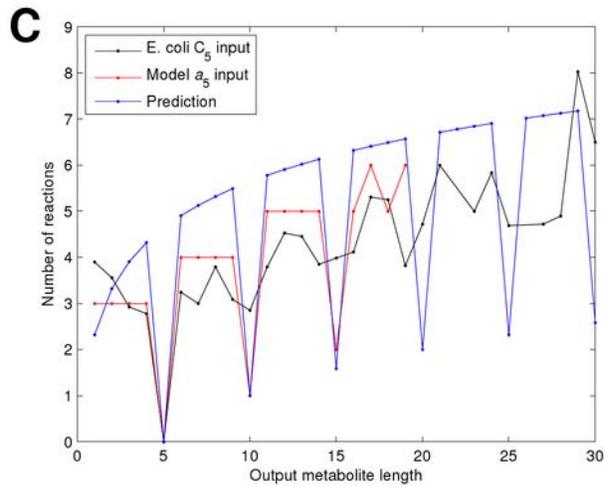

# Figure 4

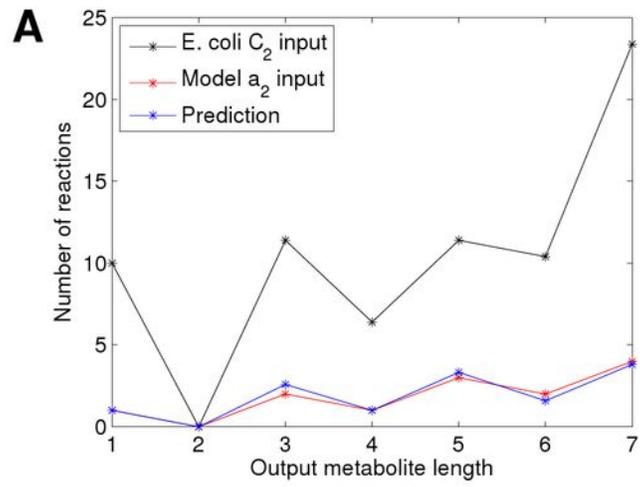

**A**

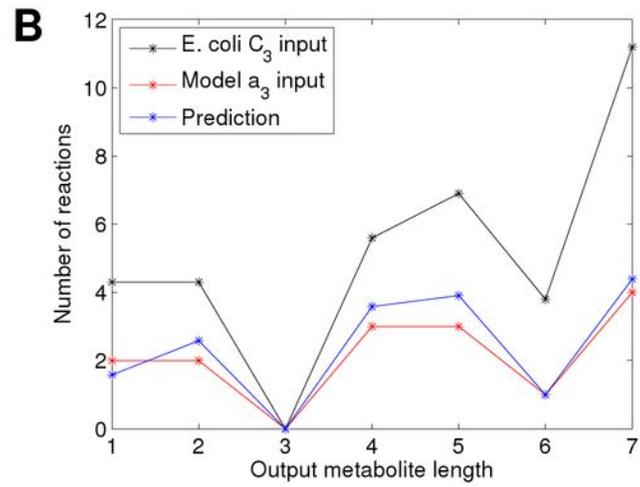

**B**

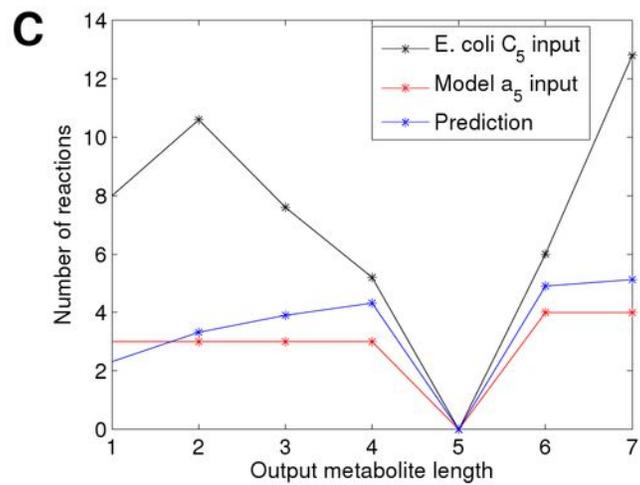

**C**

# Figure 5

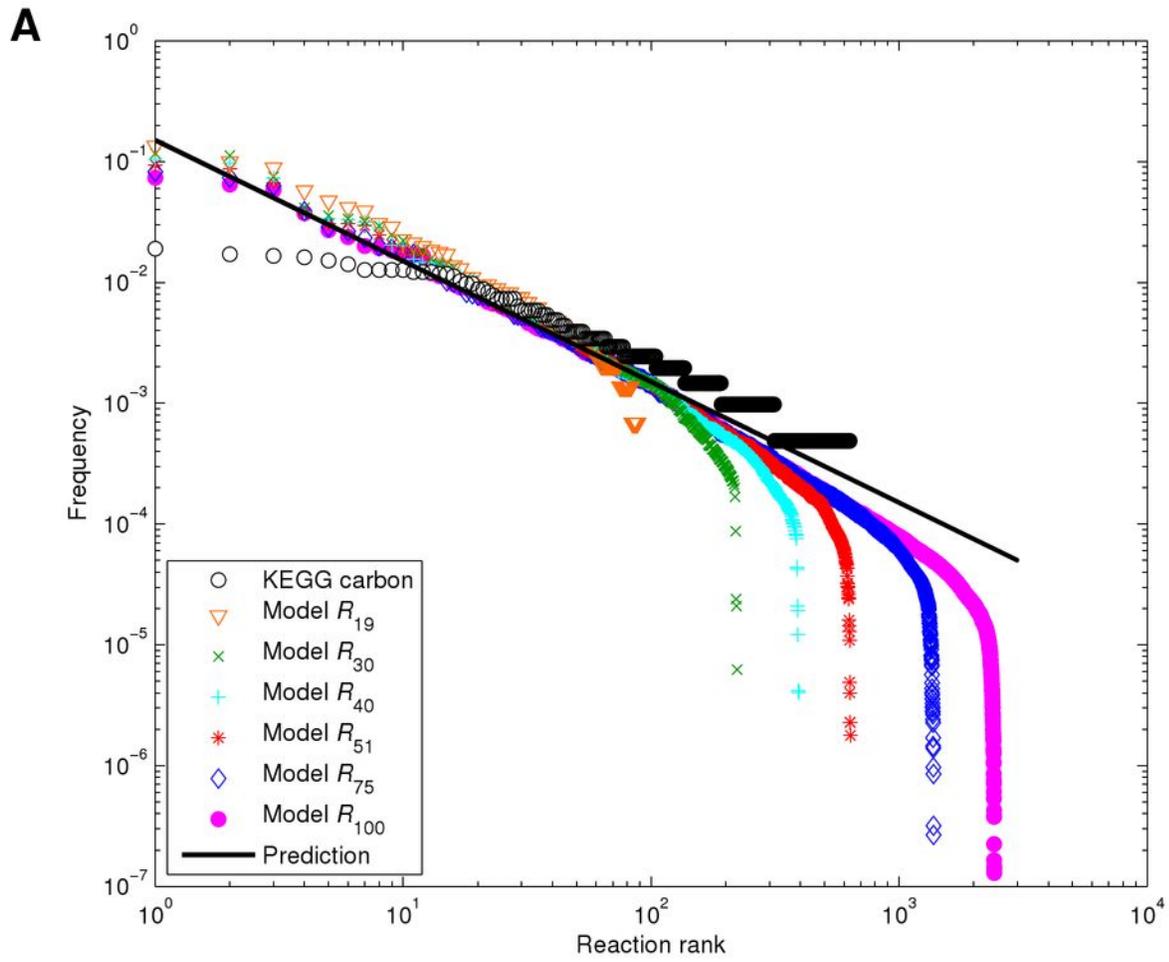

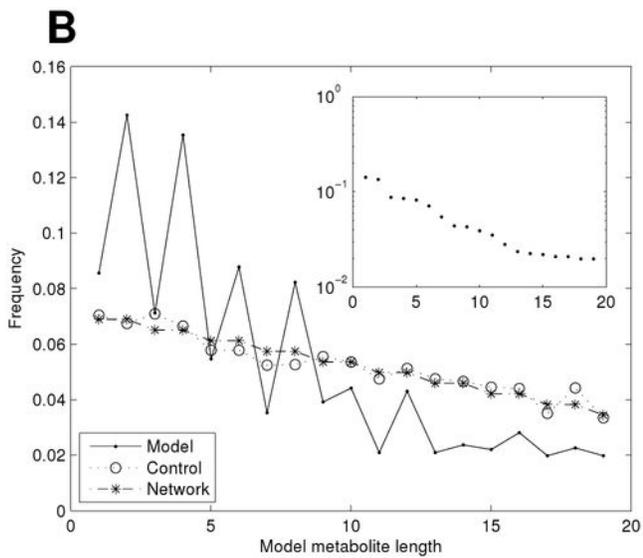

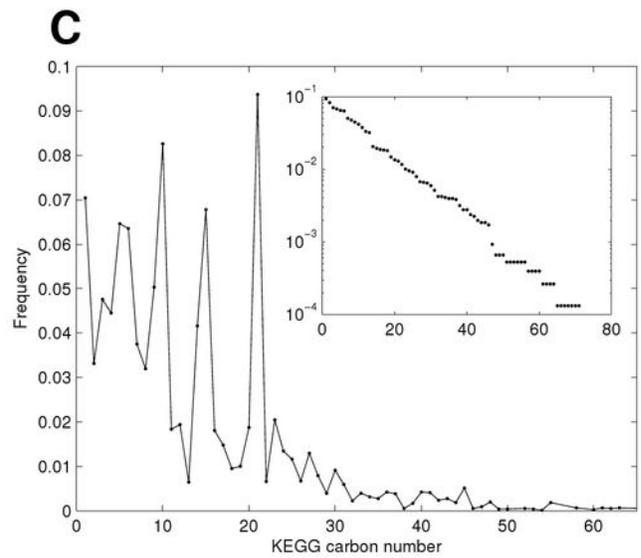